\documentclass[10pt,twocolumn,nofootinbib,aps,prl]{revtex4-1}
\usepackage{amsmath,amssymb,bm,mathrsfs}
\usepackage{mathtools,array,slashed}
\usepackage{graphicx,graphics,subfigure}
\usepackage{ctable,multirow}
\usepackage{tabularx}
\usepackage[colorlinks=true,linkcolor=blue,citecolor=blue,urlcolor=blue]{hyperref}
\usepackage{color,xcolor}

\def\be{\begin{equation}}
\def\ee{\end{equation}}
\def\bea{\begin{eqnarray}}
\def\eea{\end{eqnarray}}
\def\ba{\begin{aligned}}
\def\ea{\end{aligned}}
\def\nn{\nonumber}
\def\p{\partial}

\begin{document}


\title{Novel topological classes in black hole thermodynamics}

\author{Di Wu$^1$}
\email{Contact author: wdcwnu@163.com}

\author{Wentao Liu$^2$}

\author{Shuang-Qing Wu$^1$}
\email{Contact author: sqwu@cwnu.edu.cn}

\author{Robert B. Mann$^3$}
\email{Contact author: rbmann@uwaterloo.ca}

\affiliation{$^{1}$School of Physics and Astronomy, China West Normal University,
Nanchong, Sichuan 637002, People's Republic of China \\
$^{2}$Department of Physics, Key Laboratory of Low Dimensional Quantum Structures and
Quantum Control of Ministry of Education, and Synergetic Innovation Center for Quantum
Effects and Applications, Hunan Normal University, Changsha, Hunan 410081, People's
Republic of China \\
$^{3}$Department of Physics \& Astronomy, University of Waterloo, Waterloo,
Ont. Canada N2L 3G1}

\date{\today}

\begin{abstract}
By viewing black hole solutions as topological defects in thermodynamic parameter space, we
unveil a novel topological class and two new topological subclasses, respectively, denoted as
$W^{0-\leftrightarrow 1+}$, $\overline{W}^{1+}$, and $\hat{W}^{1+}$, that extend beyond the
four established categories proposed by Wei \textit{et al.} [\href{http://dx.doi.org/10.1103
/PhysRevD.110.L081501}{Phys. Rev. D \textbf{110}, L081501 (2024)}]. Within the newly identified
class and these two novel subclasses, the innermost small black hole states exhibit a distinct
sequence of unstable, stable, and stable behaviors, while the outermost large black hole states
display a uniform pattern of stable behaviors. These classifications indicate thermodynamic
properties both in the low and high Hawking temperature regimes that are strikingly different
from the previously known four topological classes. In particular, we demonstrate that the
static charged AdS black holes in gauged supergravity exhibit an intricate thermodynamic
evolution that is notably distinct from that of  the Reissner-Nordstr\"{o}m anti-de Sitter
(RN-AdS) black hole. From a topological perspective, we emphasize the advantages and potential
of investigating thermodynamic phase transitions in these black hole spacetimes, an area that
has been rarely explored in the previous research. Our findings not only enrich and sharpen
the framework of topological classifications in black hole thermodynamics, but also represent
a significant stride toward unraveling the fundamental nature of black holes and gravity.
\end{abstract}

\maketitle

\textit{Introduction}.---Black holes serve as a crucial testing ground for quantum gravity
theories, stemming from the discovery that they exhibit thermal properties, with entropy
proportional to the area of their event horizons \cite{PRD7-2333,CMP43-199}. Over the past
two decades, our understanding of black hole mechanics has expanded to include concepts of
pressure and volume \cite{PRD52-4569,CQG17-399,CPL23-1096,CQG26-195011,PRD84-024037}. This
area, commonly referred to as black hole chemistry \cite{CQG34-063001}, has provided new
insights into gravitational phase transitions \cite{CMP87-577,JHEP0712033,PRD88-121502,
PRL118-021301} and entropy bounds \cite{PRL130-121501,PRL131-241401,2406.17860}, with a
holographic interpretation now becoming established that includes heat engines \cite{CQG31-205002},
complexity \cite{PRL126-101601}, the significance of the central charge \cite{PRL127-091301,
PRD105-106014,JHEP0822174,CQG39-075019,PRL130-181401,JHEP0623105,JHEP0823142}, and its origins
in higher dimensions \cite{PRL130-161501}.

Although there has been significant progress in recent years, exploring the universal properties
of black hole thermodynamics continues to be a complex undertaking. Recent developments suggest
that topology can sheds new light on addressing this issue \cite{PRL129-191101,PRD110-L081501,
PRD105-104003} by considering black hole solutions as topological defects in the thermodynamic
parameter space. Initially, such defects were classified into three categories based on their
different topological numbers \cite{PRL129-191101}. This approach was later refined in Ref.
\cite{PRD110-L081501}, categorizing black hole solutions into four more general classes based
on their thermodynamic asymptotic behaviors. For more examples of the latest representative
developments, see Refs. \cite{PRD107-024024,PRD107-064023,PRD107-084002,EPJC83-365,EPJC83-589,
PRD108-084041,PLB856-138919,PDU46-101617,2409.11666,PLB860-139163}. Furthermore, the topology
of thermodynamics and phase transition for certain AdS black holes have been studied in Refs.
\cite{JHEP0623115,EPJC84-43,AP465-169679,2501.00955} in the context of bulk, mixed bulk/boundary,
and conformal field theory (CFT) thermodynamics by using different holographic dictionaries for the dual CFT. In particular, a residue method was adopted in \cite{JHEP0623115} to study the
topological properties of the phase transitions of charged AdS$_4$ black holes and showed that
the bulk and boundary thermodynamics are topologically equivalent for both criticality and
first-order phase transition in the canonical ensembles. On the other hand, Ref. \cite{AP465-169679} showed that a Born-Infeld-AdS black hole and its dual CFT share the identical topology, so there is a parallel transition in the dual CFT corresponding to the topological transition that occurs in the bulk.

However the thermodynamic topological classification of black hole solutions into four distinct
classes, as proposed in Ref. \cite{PRD110-L081501}, does not cover all black hole solutions.
For example, because of the novel temperature-dependent thermodynamic topological phase
transition occurring in the four-dimensional static charged AdS black hole of the
Einstein-Maxwell-dilaton-axion (EMDA) gauged supergravity theory and the four-dimensional
static two-charge AdS black hole, as well as the five-dimensional static charged AdS black hole
within the Kaluza-Klein (K-K) gauged supergravity theory \cite{JHEP0624213}, these three black
hole solutions cannot be included into any of the existing four topological classes. In addition,
the dyonic AdS black hole \cite{2405.07525} exhibits certain differences from the existing four
topological classes, making it unsuitable for inclusion into these categories.

Therefore, a natural question arises as to whether new topological classes (or subclasses)
exist to accommodate these situations. In this Letter, our findings indicate the existence
of one novel topological class and two new topological subclasses beyond the four known classes.
In addition, we aim to show that static charged AdS black holes in gauged supergravity, compared
to the RN-AdS black hole, display more distinct and intricate thermodynamic evolution. We also
highlight the considerable advantages and potential of investigating thermodynamic phase
transitions in static charged AdS black holes within gauged supergravity theories from a
topological viewpoint, an area that remains largely unexplored in current research.

The topological properties of black hole thermodynamics fundamentally depend on the generalized
off-shell free energy \cite{PRD33-2092}. In this approach, a black hole with mass $M$ and
entropy $S$ is enclosed in a cavity at a fixed temperature $1/\tau$, leading to a generalized
free energy $\mathcal{F} = M -S/\tau$, which can also be derived from the gravitational path
integral \cite{PRD106-106015}. The free energy reduces to an on-shell quantity only when $\tau
= \beta = 1/T$, where $T$ is the Hawking temperature of the black hole. By introducing an
additional parameter $\Theta \in (0, \pi)$, a two-component vector field can be defined via
\be
\phi = \big(\phi^{r_h}, \phi^\Theta\big) = \Big(\frac{\p\mathcal{F}}{\p\, r_h},
 -\frac{\cos\Theta}{\sin^2\Theta}\Big) \, .
\ee
The condition $\phi^{r_h} = 0$ identifies black hole states as zero points (or defects) of the
vector field. Utilizing Duan's $\phi$-mapping topological current theory \cite{SS9-1072}, each
zero point or black hole state is then assigned with a topological charge, known as the winding
number $w$ \cite{PRL129-191101}.

Within this framework, from a local perspective, black hole states that are locally stable have
a positive local topological number $w = +1$, while locally unstable states are characterized
by a negative local topological number $w = -1$. Furthermore, for a given class of black holes,
the sum of all local topological numbers $w_i$ yields the global topological number $W$. Therefore,
from a global perspective, black holes can be topologically classified according to their global
topological number $W$.

\begin{figure}[t]
\subfigure[~{Typical case: Schwarzschild black hole.}]
{\label{W1-}
\includegraphics[width=0.22\textwidth]{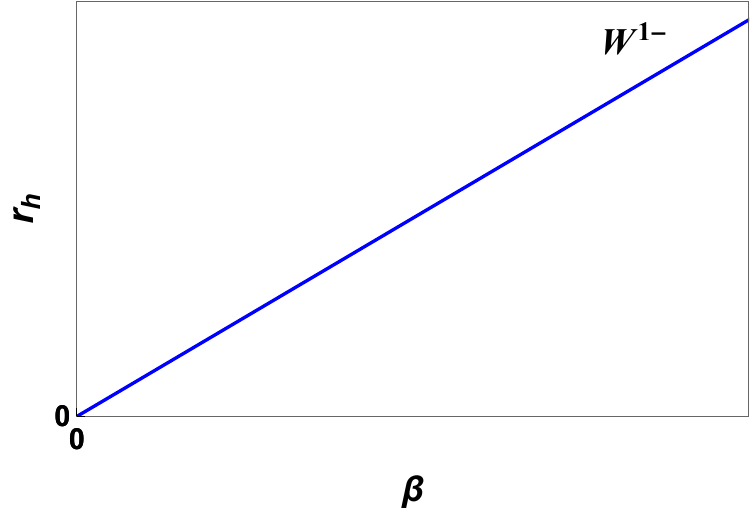}}
\subfigure[~{Typical case: RN black hole.}]
{\label{W0+}
\includegraphics[width=0.22\textwidth]{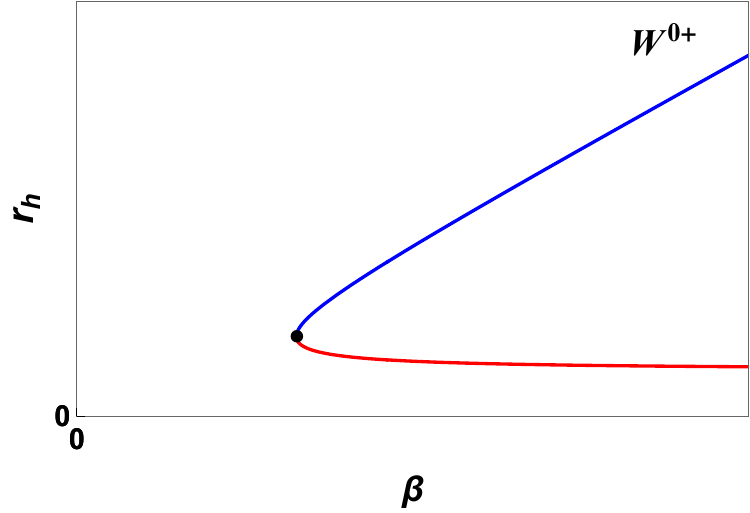}}
\subfigure[~{Typical case: Schwarzschild-AdS black hole.}]
{\label{W0-}
\includegraphics[width=0.22\textwidth]{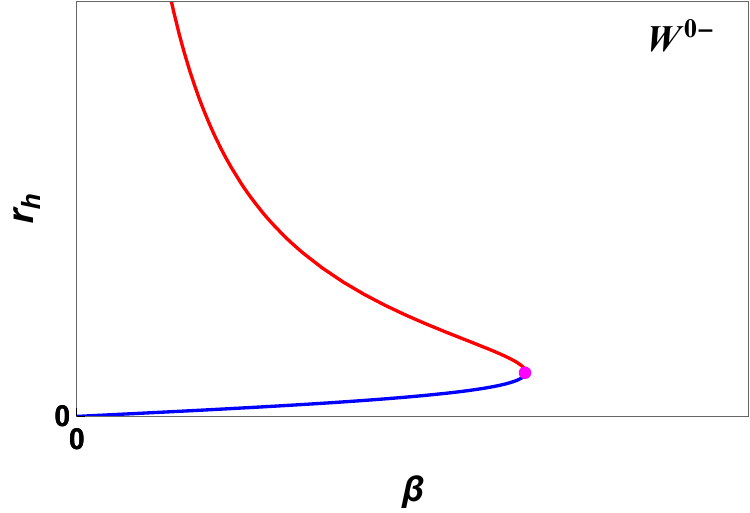}}
\subfigure[~{Typical case: RN-AdS black hole.}]
{\label{W1+}
\includegraphics[width=0.22\textwidth]{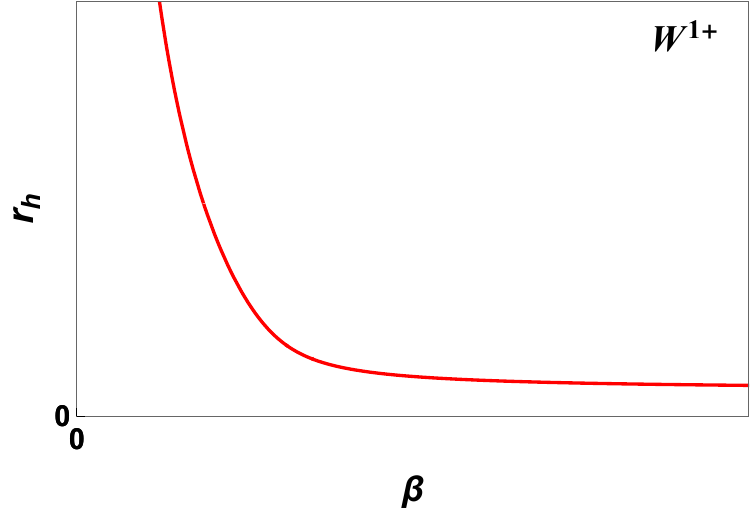}}
\caption{The zero points of $\phi^{r_h}$ are shown in the $r_h-\beta$ plane
for $W^{1-}$, $W^{0+}$, $W^{0-}$, $W^{1+}$ classes, respectively. The blue
line corresponds to a thermodynamically unstable black hole branch with $w
= -1$, while the red line corresponds to a thermodynamically stable black
hole branch with $w = 1$. The black dot represents the generation point (GP)
within the degenerate point (DP), and the pink dot represents the annihilation
point (AP) within the DP. \label{fig1}}
\end{figure}

\textit{Four known topological classes}.---Let us start with a brief review of the four known
thermodynamic topological classes previously specified in \cite{PRD110-L081501}
\be
W^{1-} \, , \quad W^{0+} \, , \quad W^{0-} \, , \quad W^{1+} \, ,
\ee
which correspond to the following asymptotic behaviors of the inverse temperature $\beta(r_h)$:
\bea
&W^{1-}&:~  \beta(r_{m})=0\, ,\quad\;\; \beta(\infty)=\infty \, ,  \\
&W^{0+}&:~  \beta(r_{m})=\infty \, ,\quad \beta(\infty)=\infty \, ,   \\
&W^{0-}&:~  \beta(r_{m})=0\, ,\quad\;\; \beta(\infty)=0 \, , \label{b0-} \\
&W^{1+}&:~  \beta(r_{m})=\infty \, ,\quad \beta(\infty)=0 \, , \label{b1+}
\eea
where $r_{m}$ is the minimal radius of the black hole event horizon, which may or may not
vanish. For instance, a RN black hole with a fixed charge $Q$ satisfies $r_{m} = M = Q = r_e$
in the extremal case, while for a Schwarzschild black hole, $r_{m}$ vanishes. Table II in Ref.
\cite{PRD110-L081501} and Fig. \ref{fig1} below provide a comprehensive summary of the
properties of the four known topological classes, facilitating a better understanding of
their characteristics.

\textit{One novel topological class}.---Recently we observed \cite{JHEP0624213} a novel
temperature-dependent thermodynamic topological phase transition occurring in the
four-dimensional static charged AdS black hole within the EMDA gauged supergravity theory,
and the four-dimensional static two-charge AdS black hole, as well as the five-dimensional
static charged AdS black hole within the K-K gauged supergravity theory.

We will take the four-dimensional static two-charge AdS black hole as an example to construct
a new classification that includes black hole solutions exhibiting temperature-dependent
thermodynamic topological phase transitions, representing the three similar cases. The metric,
Abelian gauge potentials, scalar fields, and thermodynamic quantities of the four-dimensional
static two-charge AdS black hole are given in Eqs. (\ref{4d}) and (\ref{therm}) in \textit{
Appendix A}.

\begin{figure}[t]
\centering
\includegraphics[width=0.25\textwidth]{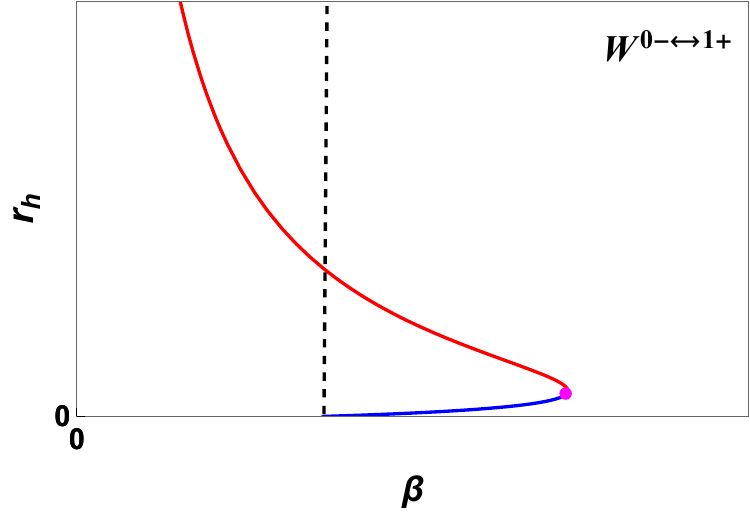}
\caption{The zero points of $\phi^{r_h}$ are shown in the $r_h-\beta$ plane
for the novel topological class $W^{0-\leftrightarrow 1+}$. A typical example
in this category is the four-dimensional static two-charge AdS black hole when
$q_1 < q_{1c} = 3/(8\pi Pq_2)$, where temperature-dependent thermodynamic
topological phase transitions occur, and an annihilation point (pink dot)
is displayed. Red (blue) curves denote stable (unstable) branches. \label{fig2}}
\end{figure}

Consider first the case in which the four electric charge parameters are set to $q_1 \ne q_2
\ne 0$. For a fixed electric charge parameter $q_2$ and a fixed pressure $P$, when $q_1 <
q_{1c} = 3/(8\pi Pq_2)$, the topological number of the four-dimensional static two-charge
AdS black hole is either $W = 0$ (at cold temperatures) or $W = 1$ (at high temperatures),
exhibited in Fig. \ref{fig2}. It is straightforward to obtain
\be\label{nb}
\beta(r_m) = {\rm fixed ~\, temperature} \, , \quad \beta(\infty) = 0
\ee
for the asymptotic behavior of the inverse temperature $\beta(r_h)$ of this new topological
class.

Next, we analyze the systematic ordering for the new topological class. There is at least
one black hole state with negative heat capacity and a winding number with $-1$, and one
black hole state with positive heat capacity and a winding number with $+1$. If additional
black hole states are present, they must emerge in pairs. As the signs of the heat capacities
alternate with increasing $r_h$, the smallest state corresponds to a thermodynamically
unstable black hole, while the largest state corresponds to a thermodynamically stable one.
The winding numbers associated with the zero points follow the sequence $[-, (+, -), ..., +]$,
where the ellipsis represents pairs of $(-, +)$ winding numbers. For simplicity, this novel
topological class can be labeled as $[-, +]$ based upon the signs of the innermost and
outermost winding numbers.

Turning to the asymptotic thermodynamic  behavior of this new topological class, in the
low-temperature limit, $\beta \to \infty$, no black hole state is present. At the high-temperature
limit, $\beta \to 0$, the system features a stable large black hole. On the base of its unique
thermodynamic behavior and following the consistent naming convention from small black hole
states to large black hole states, we will name this novel topological classification:
\be
W^{0-\leftrightarrow 1+}
\ee
indicating that black hole solutions belonging to this classification display thermodynamic
topological phase transitions.

Last but not least, let us revisit the unstable black hole branch shown in Fig. \ref{fig2}
(depicted by the blue line) to examine its distinctive thermodynamic behavior, which cannot
be observed in the RN-AdS black hole. Thermal instability typically implies that a black hole
has a negative heat capacity, implying that as the black hole loses energy, its temperature
increases. This negative heat capacity effect renders the black hole unstable under energy
absorption or emission. A static charged AdS black hole in gauged supergravity theory
exhibiting thermal instability at higher temperatures may evolve along this potential
path: at elevated temperatures, the black hole radiation intensifies, potentially leading
to a rapid loss of mass and charge, accelerating the evaporation process. If both mass and
charge deplete swiftly, the system may evolve into a horizonless, charged AdS spacetime.
During this process, as the black hole mass decreases, its temperature rises further due
to the negative heat capacity, enhancing the rate of charge and energy emission. This
positive feedback mechanism accelerates the evaporation of the black hole, leading to
the disappearance of its event horizon. Note that in the horizonless, charged AdS spacetime,
the electric charge typically is manifest as a far-field effect, predominantly concentrated
at the boundary of the spacetime or distributed via the electric field to large distances.

\begin{figure}[h]
\centering
\includegraphics[width=0.25\textwidth]{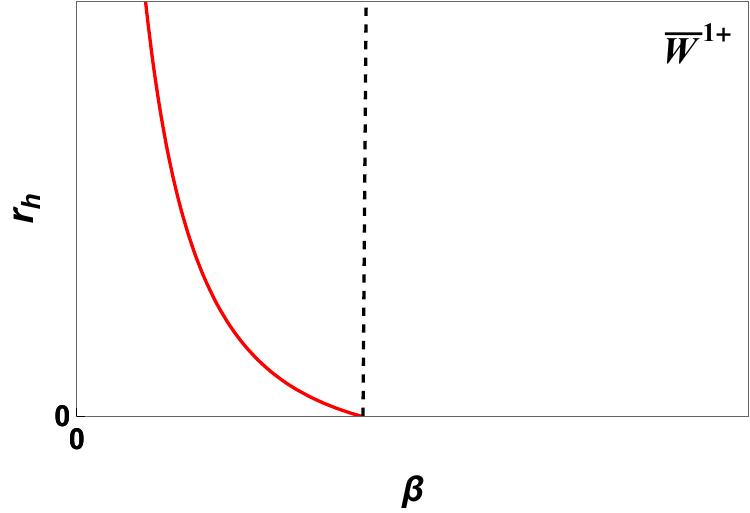}
\caption{The zero points of $\phi^{r_h}$ are shown in the $r_h-\beta$ plane for
the first new topological subclass $\overline{W}^{1+}$, which has the topological
number $W = 1$. A typical example in this category is the four-dimensional static
two-charge AdS black hole when $q_1 \ge q_{1c} = 3/(8\pi Pq_2)$. \label{fig3}}
\end{figure}

\textit{Two new topological subclasses}.---We now describe two new topological subclasses
using two examples:

The first is the four-dimensional static two-charge AdS black hole when $q_1 \ge q_{1c} =
3/(8\pi Pq_2)$. Its topological number is $W = 1$, and its thermodynamic behavior is shown
in Fig. \ref{fig3}. Similar thermodynamic behavior can also be found in the four-dimensional
static charged AdS black hole in the EMDA gauged supergravity for $q \ge \sqrt{6}/(4\sqrt{\pi
P})$ and the five-dimensional static charged AdS black hole in the K-K gauged supergravity
for $q \ge 3/(4\pi P)$ \cite{JHEP0624213}. By comparing Fig. \ref{fig3} and Fig. \ref{W1+},
it is evident that their thermodynamic behaviors are significantly different. However, since
both share the same topological number of one, this should represent a new topological subclass.
The asymptotic behavior of the inverse temperature $\beta(r_h)$ for this new topological subclass
is the same one as that shown in Eq. (\ref{nb}), but the value of the fixed temperature may be
different. This novel topological subclass can be labeled as $[+, +]$ due to the signs of the
innermost and outermost winding numbers. In the high-temperature limit, $\beta \to 0$, the
system exhibits a stable large black hole state, which is consistent with that of the $W^{1+}$
class. However, in the low-temperature limit, $\beta \to \infty$, no black hole state exists,
distinguishing it from the $W^{1+}$ class. We thus label this new topological subclass as
$\overline{W}^{1+}$.

\begin{figure}[h]
\centering
\includegraphics[width=0.25\textwidth]{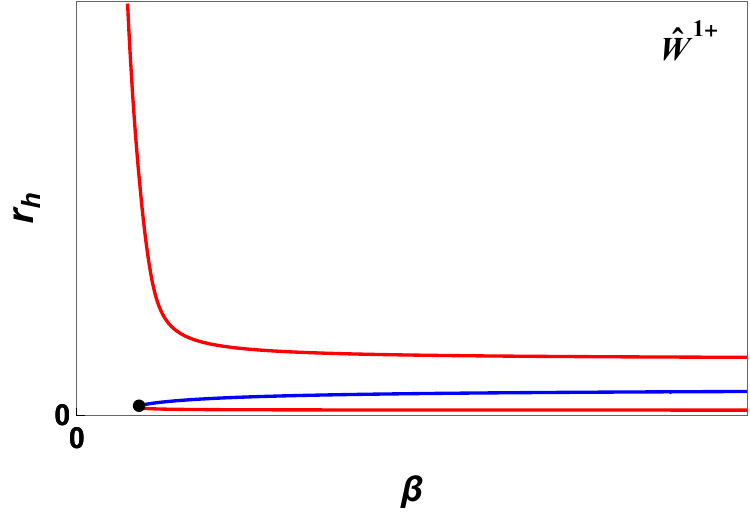}
\caption{The zero points of $\phi^{r_h}$ are shown in the $r_h-\beta$ plane
for the third new topological subclass $\hat{W}^{1+}$, which has the topological
number $W = 1$. A typical example in this category is the four-dimensional dyonic
AdS black hole. The black dot denotes a generation point. Red (blue) curves denote
stable (unstable) branches. \label{fig4}}
\end{figure}

Our second example is the four-dimensional dyonic AdS black hole in Refs. \cite{2405.07525,
SCPMA63-240411}. Its metric and thermodynamic quantities are displayed by Eqs. (\ref{4dd})
and (\ref{Therm}) in \textit{Appendix B}. Its topological number is $W = 1$, and its thermodynamic
behavior is illustrated in Fig. \ref{fig4}. A comparison of Figs. \ref{fig4} with \ref{fig3}
and \ref{W1+} reveals that, despite that all of them have a topological number of one, their
thermodynamic behaviors are distinct, indicating a second new topological subclass, which we
will designate as $\hat{W}^{1+}$. The asymptotic behavior of the inverse temperature $\beta(r_h)$
aligns with the pattern presented in Eq. (\ref{b1+}), and we can label this subclass $[+,+]$
according to the signs of its innermost and outermost winding numbers. In the high-temperature
limit, $\beta\to 0$, the system exhibits a stable large black hole state, paralleling the
behavior of the $W^{1+}$ and $\overline{W}^{1+}$ classes. However, in the low-temperature limit,
$\beta\to\infty$, it features two stable small black hole states and one unstable small black
hole state, which sets it apart from the $W^{1+}$ and $\overline{W}^{1+}$ classes.

\textit{Holographic dual CFT side}.---Through the holographic dictionary \cite{PRL130-181401}:
$E = M/\omega$, $\tilde{T} = T/\omega$, and $\tilde{S} = S$, one can define the free energy for
a canonical ensemble in the dual boundary CFT: $\tilde{\mathcal{F}} = E -\tilde{S}/\tilde{\tau}
= \mathcal{F}/\omega$, where the Euclidean period time in the dual boundary theory is related to
that in the bulk via $\tilde{\tau} = \omega\tau$ by virtue of the above dictionary, and $\omega =
\mathcal{R}/l$ is an arbitrary conformal factor of the boundary AdS metric, in which $\mathcal{R}$
is the boundary curvature radius, which is different from the bulk curvature radius $l$. Clearly,
one can reasonably assume that the boundary conformal factor should be independent of the horizon
radius, that is, $\p\omega/{\p\, r_h} = 0$. Then similarly to the bulk case, one can define a
two-component vector field $\tilde{\phi}$ in the boundary CFT with its radial component being
$\tilde{\phi}^{r_h} = \phi^{r_h}/\omega$ that only zooms in or zooms out the size of $\phi^{r_h}$.
This implies that the orientations of the vector fields both in the bulk and in the boundary
theories are coincident, which leads us to draw a \emph{universal} conclusion that the topology
of boundary CFT thermodynamics shares the same one of the bulk thermodynamics in the sense of
canonical ensembles. This would be an very expected and promising result in that it reflects
the gravity-gauge duality in terms of topology, which depicts the global character of the
geometry. Otherwise, it would be rather odd and unappreciated.

\textit{Conclusions and discussions}.---We have introduced a new thermodynamic topological class
$W^{0-\leftrightarrow 1+}$ and two novel subclasses $\overline{W}^{1+}$, and $\hat{W}^{1+}$
that extend beyond the four previously established categories \cite{PRD110-L081501}. For each
new topological (sub)class, we analyzed its thermodynamic behavior in both the small and large
black hole limits, as well as in the low and high temperature limits. A concise overview of these
results is presented in Table \ref{TableI}. Our results emphasize the significance and potential
of investigating the thermodynamic phase transition behavior of static charged AdS black holes in
gauged supergravity theories, whose thermodynamic behavior is far richer than the corresponding
RN-AdS case.

In summary, our findings highlight both the necessity and great potential of studying the
thermodynamic phase transition behavior of static charged AdS black holes in gauged supergravity
theories, though we have just grasped a fraction of these insights from the perspective of
thermodynamic topology. More generally, we have enhanced and refined the framework of topological
classifications in black hole thermodynamics, while also marking a significant step toward
uncovering the fundamental nature of black holes and gravity.

\onecolumngrid
\begin{center}
\begin{table}[b]
\begin{tabular}{c|c|c|c|c|c|c}\hline\hline
Classes  & Innermost  & Outermost  & Low $T$ ($\beta\to\infty$)
 & High $T$ ($\beta\to 0$)  & DP  & $W$ \\ \hline
$W^{0-\leftrightarrow 1+}$  & unstable  & stable  & no
 & stable large  & one more AP & $0$ or $+1$ \\
$\overline{W}^{1+}$  & stable  & stable  & no
 & stable large  & in pairs & $+1$ \\
$\hat{W}^{1+}$  & stable  & stable  & unstable small
 +two stable small  & stable large  & one more GP  & $+1$ \\
\hline\hline
\end{tabular}
\caption{Thermodynamical properties of the black hole states for the one novel
topological class and two new topological subclasses of $W^{0-\leftrightarrow 1+}$,
$\overline{W}^{1+}$, and $\hat{W}^{1+}$, respectively.} \label{TableI}
\end{table}
\end{center}

\begin{center}
\begin{table}[t]
\begin{tabular}{c|c|c|c|c|c|c}\hline\hline
Classes  & Innermost  & Outermost  & Low $T$ ($\beta\to\infty$)
 & High $T$ ($\beta\to 0$)  & DP  & $W$ \\ \hline
$W^{0+\leftrightarrow 1-}$  & stable  & unstable  & unstable large
 & no  & one more GP  & $0$ or $-1$ \\
$\overline{W}^{1-}$  & unstable  & unstable  & unstable large
 & no  & in pairs  & $-1$ \\
$\hat{W}^{1-}$  & unstable  & unstable  & unstable large
 & stable small +two unstable small  & one more AP  & $-1$ \\
\hline\hline
\end{tabular}
\caption{Thermodynamical properties of the black hole states for the possible
new topological (sub)classes of $W^{0+\leftrightarrow 1-}$, $\overline{W}^{1-}$,
and $\hat{W}^{1-}$, respectively.}\label{TableII}
\end{table}
\end{center}
\twocolumngrid

Our results also raise an interesting question: Are there any new topological classes or subclasses
yet to be identified? By examining the symmetry of functions ($r_h, \beta$), one can conjecture
there might still exist another three new thermodynamic topological (sub)classes based upon
the three novel (sub)classes we have supplemented, although it remains unclear whether their
corresponding black hole solutions actually exist or not, which needs to be verified in future
research. These possible new (sub)classes are listed as follows:
\be
W^{0+\leftrightarrow 1-} \, , \quad \overline{W}^{1-} \, , \quad \hat{W}^{1-} \, ,
\ee
which are aligned with the asymptotic behaviors of the inverse temperature $\beta(r_h)$:
\bea
&W^{0+\leftrightarrow 1-}&:~  \beta(r_{m})= {\rm fixed ~\, temperature} \, ,\quad
 \beta(\infty)=\infty \, ,   \nn \\
&\overline{W}^{1-}&:~  \beta(r_{m})= {\rm fixed ~\, temperature} \, ,\quad
 \beta(\infty)=\infty \, , \qquad \nn \\
&\hat{W}^{1-}&:~  \beta(r_{m})=0 \, ,\qquad\qquad\qquad\qquad~~
 \beta(\infty)= \infty \, . \nn
\eea

\begin{figure}[t]
\subfigure[]
{\label{W1-}
\includegraphics[width=0.15\textwidth]{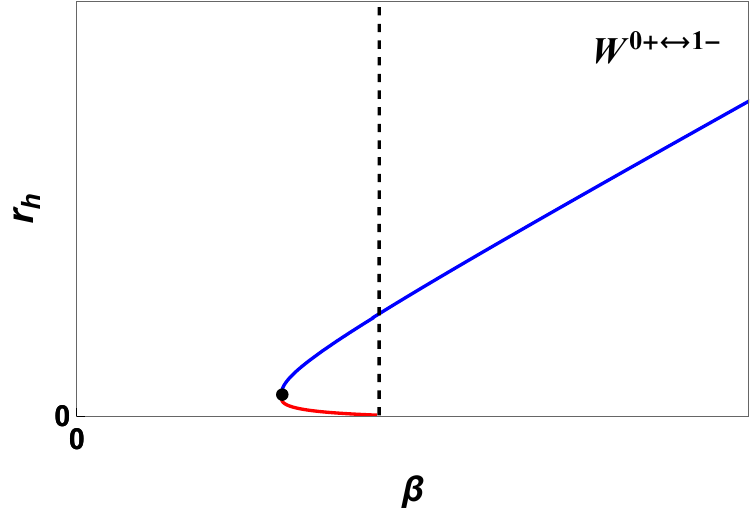}}
\subfigure[]
{\label{W0+}
\includegraphics[width=0.15\textwidth]{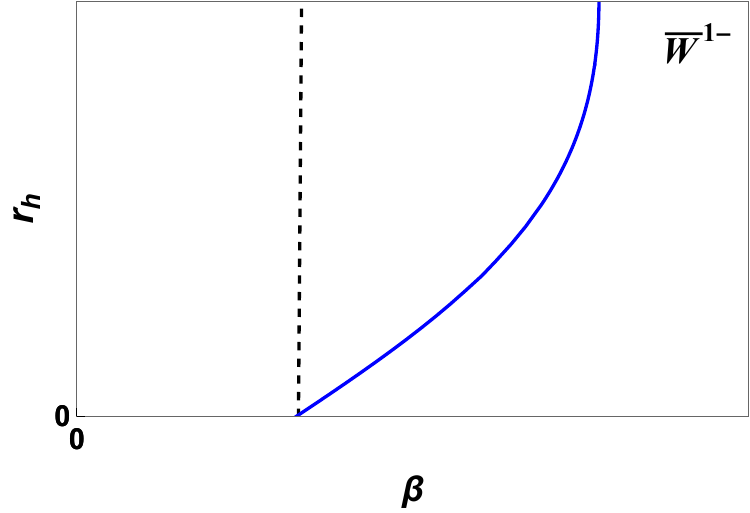}}
\subfigure[]
{\label{W0-}
\includegraphics[width=0.15\textwidth]{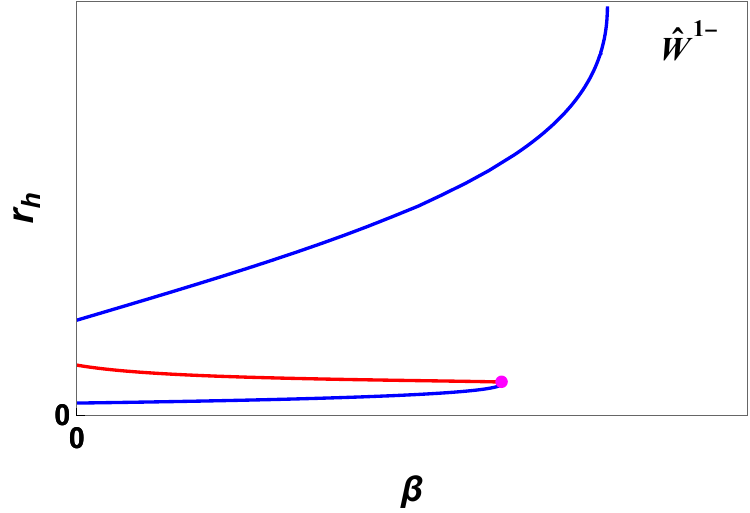}}
\caption{The schematic diagram of the zero points of $\phi^{r_h}$ is shown
in the $r_h-\beta$ plane for the possible new topological (sub)classes: (a)
$W^{0+\leftrightarrow 1-}$; (b) $\overline{W}^{1-}$; (c) $\hat{W}^{1-}$.
\label{fig5}}
\end{figure}

We have also provided a schematic diagram in Fig. \ref{fig5} and summarized in Table \ref{TableII} the thermodynamical properties of the three potentially existing topological (sub)classes of $W^{0+\leftrightarrow 1-}$, $\overline{W}^{1-}$, and $\hat{W}^{1-}$, respectively. Certainly, further exploration is necessary to verify the potential existence of the corresponding black hole solutions. Perhaps we have only touched the tip of the iceberg: these and many other questions remain intriguing subjects for future investigation.

\textit{Acknowledgments}.---We are greatly indebted to the anonymous referee for his/her constructive comments to improve the presentation of this work. We also thank Prof. S.-W. Wei for very helpful discussions. This work is supported by the National Natural Science Foundation of China (NSFC) under Grants No. 12205243, No. 12375053, by the Sichuan Science and Technology Program under Grant No. 2023NSFSC1347, by the Doctoral Research Initiation Project of China West Normal University under Grant No. 21E028, by the Postgraduate Scientific Research Innovation Project of Hunan Province (CX20240531), and by the Natural Sciences and Engineering Research Council of Canada.

\medskip

\setcounter{equation}{0}
\renewcommand{\theequation}{A\arabic{equation}}

\textit{Appendix A: $D=4$ static four-charge AdS black hole in STU gauged supergravity}.---For
the general static four-charge AdS black hole in four-dimensional STU gauged supergravity,
the metric, the Abelian gauge potentials, and the scalar fields are given by \cite{NPB554-237}:
\bea\label{4d}
ds^2 &=& -\prod_{i=1}^4 H_i^{-1/2}\, f dt^2 +\prod_{i=1}^4 H_i^{1/2}\big(f^{-1}{dr^2}
 +r^2d\Omega_{2}^2 \big) \, ,\nn \\
A^i &=& \frac{\sqrt{q_i(q_i +2m)}}{2(r +q_i)}\, dt\, ,\quad
X_i = H_i^{-1}\, \prod_{j=1}^4 H_j^{1/4}\, ,
\eea
where
\be
f = 1 - \frac{2m}{r} + \frac{r^2}{l^2}\prod_{i=1}^4H_i \, , \qquad
H_i = 1 +\frac{q_i}{r} \, , \nn
\ee
in which $l$ is the AdS radius, $m$ and $q_i$ are the mass and four independent electric
charge parameters, respectively.

For the static charged AdS black hole metric described by Eq. (\ref{4d}), the most general
solution features four independent electric charge parameters. In addition, based upon the
classification of black hole solutions shown in Fig. 1 of Ref. \cite{PRD90-025029}, there
are several notable truncated supergravity solutions. For instance, when $q_1 \ne q_2 \ne 0$,
$q_3 = q_4 = 0$, this corresponds to the four-dimensional static two-charge AdS black hole
solution \cite{CQG28-175004}. When $q_1 = q_2 \ne 0$ and $q_3 = q_4 = 0$, it describes the
static charged AdS black hole solution in the EMDA gauged supergravity theory \cite{PRD102-044007,
PRD103-044014}. In addition, when $q_1 = q_2 = q_3 = q_4 \ne 0$, the solution reduces to the
familiar RN-AdS black hole after a coordinate transformation $\rho = r +q_i$, etc.

The thermodynamic quantities are given by \cite{PRD84-024037}
\bea\label{therm}
&& M = m +\frac{1}{4}\sum_{i=1}^4q_i\, , \quad Q_i= \frac{1}{2}\sqrt{q_i(q_i +2m)}\, , \nn \\
&& S = \pi\prod_{i=1}^4 (r_h +q_i)^{1/2}\, , \quad
 T = \frac{f'(r_h)}{4\pi}\prod_{i=1}^4 H_i^{-1/2}(r_h)\, , \nn \\
&& \Phi_i = \frac{\sqrt{q_i(q_i +2m)}}{2(r_h +q_i)}\, , \quad  P = \frac{3}{8\pi l^2} \, , \nn \\
&& V = \frac{\pi r_h^3}{3}\prod_{i=1}^4 H_i(r_h)\sum_{j=1}^4\frac{1}{H_j(r_h)} \, .
\eea

It is easy to verify that these thermodynamic quantities simultaneously obey the first law and
the Bekenstein-Smarr mass formula
\bea
dM &=& TdS +\sum_{i=1}^4\Phi_idQ_i +VdP \, , \\
M &=& 2TS +\sum_{i=1}^4\Phi_iQ_i -2VP \, .
\eea

\setcounter{equation}{0}
\renewcommand{\theequation}{B\arabic{equation}}

\textit{Appendix B: $D=4$ dyonic AdS black hole}.---In four dimensions, the Lagrangian of
dyonic AdS black holes can be expressed as \cite{SCPMA63-240411}
\be
\mathcal{L} = \sqrt{-g}\Big\{R -2\Lambda -F^2
 -\alpha\big[(F^2)^2 -2F^{(4)}\big]\Big\} \, ,
\ee
where the standard Maxwell Lagrangian and quasi-topological electromagnetism are defined,
respectively, via $F^2 = -F^{\mu}_{~\nu}F^{\nu}_{~\mu}$, $F^{(4)} = F^{\mu}_{~\nu} F^{\nu}_{~\rho}
F^{\rho}_{~\sigma}F^{\sigma}_{~\mu}$, and the Faraday-Maxwell tensor reads $F_{\mu\nu}= \p_{\mu}
A_{\nu} -\p_{\nu}A_{\mu}$ with $A_{\mu}$ being the Abelian gauge potential. In the above, $\Lambda$
denotes the cosmological constant, the coupling constant $\alpha_2 = \alpha$ has the dimension of
$[\text{length}]^{2}$. (Note that here we have set $\alpha_1 = 1$).
	
The black hole solution in the static and spherically symmetric spacetime background has
been obtained by \cite{SCPMA63-240411}
\bea\label{4dd}
ds^2 &=& -f(r) dt^2 +\frac{1}{f(r)} dr^2 +r^2\big(d\theta^2 +\sin^2\theta\, d\phi^2\big)\, , \qquad \\
f(r) &=& 1 -\frac{2M}{r} -\frac{\Lambda}{3}r^2 +\frac{p^2}{r^2} 
 +\frac{q^2}{r^2}\, {_2}F_1\Big[\frac{1}{4}, 1; \frac{5}{4};
 -\frac{4\alpha\,p^2}{r^4}\Big] \, , \nn
\eea
where $p$ and $q$ are the electric charge and magnetic charge parameters, respectively.

The thermodynamic quantities are \cite{2405.07525,SCPMA63-240411}
\bea
&& M =\frac{r_h^2 +p^2}{2r_h} +\frac{4\pi}{3}P\, r_h^3 +\frac{q^2}{2r_h}\,{_2}F_1
 \Big[\frac{1}{4}, 1; \frac{5}{4}; -\frac{4\alpha\,p^2}{r_h^4}\Big] \, , \nn \\
&& Q_e = q \, , \quad Q_m = p \, , \quad  S = \pi r_h^2 \, , \quad
 P = -\frac{\Lambda}{8\pi} \, , \nn \\
&& T = \frac{r_h^2 -p^2}{4\pi\,r_h^3} +2P r_h
 -\frac{q^2r_h}{4\pi\big(r_h^4 +4\alpha\,p^2\big)}\, , \nn
\eea
\bea\label{Therm}
&& \Phi_e = \frac{q}{r_h}\, {_2}F_1\Big[\frac{1}{4}, 1; \frac{5}{4};
 -\frac{4\alpha\,p^2}{r_h^4}\Big] \, , \quad V = \frac{4\pi}{3}r_h^3 \, , \nn \\
&& \Phi_m = \frac{p}{r_h} +\frac{q^2r_h^3}{4p\big(r_h^4 +4\alpha\,p^2\big)} \nn \\
&&\qquad~~ -\frac{q^2}{4p\,r_h}\, {_2}F_1\Big[\frac{1}{4}, 1; \frac{5}{4};
 -\frac{4\alpha\,p^2}{r_h^4}\Big] \, ,
\eea
and have been shown to satisfy the first law
\be
dM = TdS +\Phi_e dQ_e +\Phi_m dQ_m +V dP
\ee
for fixed $\alpha$. It is also possible to verify the Smarr relation by treating $\alpha$
as a thermodynamic variable and finding its conjugate potential.

\end{document}